\documentclass [preprint, prd, aps, nofootinbib]{revtex4}

\begin{document}
\draft
\title{\large \bf An Application of Neutrix Calculus to Quantum
Field Theory}
\author{\bf Y. Jack Ng\footnote{E-mail: yjng@physics.unc.edu} and H. van Dam}
\address{Institute of Field Physics, Department of Physics and
Astronomy,\\
University of North Carolina, Chapel Hill, NC 27599-3255\\}
%\maketitle

\begin{abstract}
Neutrices are additive groups of negligible functions that do not contain any
constants except 0.  Their calculus was developed by van der Corput and Hadamard in 
connection with asymptotic series and divergent 
integrals.  We apply neutrix calculus to quantum field theory, obtaining
{\it finite} renormalizations in the loop calculations.  
For renormalizable quantum field 
theories, we recover all the usual physically observable results.  One possible
advantage of the neutrix framework is that effective field theories can be
accommodated.  Quantum gravity theories appear to be more manageable. 

%In general, quantum field theories (QFT) require regularizations and
%{\it infinite}
%renormalizations due to ultraviolet divergences in their loop calculations.
%Furthermore, perturbation series in theories like QED are not convergent
%series, but are asymptotic series.  We
%apply neutrix calculus, developed
%in connection with
%asymptotic series, to tackle divergent integrals, obtaining
%{\it finite} renormalizations for QFT.
%While none of the physically
%measurable results in renormalizable QFT is changed,
%quantum gravity is rendered more manageable
%in the neutrix framework.


\end{abstract}
\maketitle
\newpage

\section{\bf Introduction}

Quantum field theory is the proud product of quantum
mechanics and special relativity.  It is difficult to contemplate 
modern theoretical physics without it, for the very simple reason that
it is extraordinarily successful.
Coleman, a master of quantum field theory, compares the triumph of quantum
field theory to ``a glorious triumph parade, full of wonderful things'' that
make ``the spectator gasp with awe and laugh with joy.''\cite{Coleman}  
But there is a fly 
in the ointment:  in general, infinities pop up in loop corrections in
quantum field theory.  In renormalizable theories, these infinities can be
renormalized away in a process that Feynman likened to sweeping dirt 
under the rug.  Even quantum electrodynamics, arguably the most 
phenomenologically successful theory in physics, has this ``defect.''  Dirac
was quoted to have said, ``Quantum electrodynamics is almost right.''  He was
bemoaning the divergent integrals that plague loop calculations in QED.  And
it is probably not wrong to think that Schwinger put forth his source theory
approach to replace the conventional quantum field theory partly because of
the ultraviolet divergences in the latter.\cite{Schwinger}  
To put it simply, quantum field theorists are
confronted with the challenge to rid quantum field theory of infinities while 
preserving its many spectacular successes in the process.

In another development, Dyson showed that the power series in the electron 
charge $e$ for QED cannot be a convergent series.\cite{Dyson}  
His argument went roughly
as follows.  If the series is convergent, then such a series has a finite
radius of convergence in the complex plane around $e = 0$.  This means that
one can replace $e$ by $ie$ without encountering a discontinuity.  But the
physics under this replacement is actually very discontinous: electrons
would attract each other, likewise for positrons; on the other hand, 
opposite charges would repel each other so that electrons would move away 
as far as possible from positrons.  Then every physical state would be 
unstable against the spontaneous creation of large numbers of particles.
Thus the series for QED cannot be a convergent series.  Dyson went on to 
suggest that the series is an asymptotic series.  Taking his argument at 
face value, one should look for a proper tool to handle asymptotic series,
for perturbative QED in particular and perturbation calculations in 
quantum field theory in general.

Better yet, we should look for a mathematical tool that can handle 
asymptotic series in quantum field theory and rid it of infinities at the
same time.  We\cite{ngvd} have suggested that such a tool is already available
in the neutrix calculus developed by van der Corput\cite{vdC} and Hadamard.  
But instead 
of beginning with the axioms of neutrix calculus, in this paper we 
follow the path suggested to us by Prof. E.~M.~de~ Jager.  In the next 
section, we discuss the close relation between asymptotic series and
the finite parts of divergent integrals, and then introduce neutrix 
calculus to regularize divergent integrals.  To make the presentation
easy to follow, we use various examples to illustrate the approach.
In Section III, we apply
neutrix calculus to one-loop calculations in
QED and $\phi^4$ theories.  In Section IV, switching to dimensional 
regularization, we apply neutrix calculus to the pure Yang-Mills system 
and ``rederive'' asymptotic freedom in one-loop.  In both these two
sections, there is actually very little new calculation involved, but they do
serve to illustrate how neutrix calculus converts infinite renormalizations
to mathematically well-defined finite renormalizations.  As far as we can
tell, all finite physically meaningful results of renormalizable quantum
field theories are recovered in the neutrix framework.  Section V is
devoted to conclusions and a short discussion on 
the application of neutrix calculus to quantum gravity 
and effective quantum field theories in general.  For completeness,
we include two appendices to further discuss some mathematical properties of
neutrices.  In Appendix A, we consider more general Hadamard neutrices
and a more general type of asymptotic series.  To do quantum field theory in
configuration space, one multiplies operator-valued distributions of
quantum fields.  In Appendix B, we show that the use of neutrix calculus 
allows one to put these products on a mathematically sound basis. 

\bigskip

\section{\bf From asymptotic series and divergent integrals to neutrix calculus}

Neutrix calculus gives us a framework to deal with asymptotic series and 
divergent integrals.  To motivate its introduction we follow de Jager's 
suggestion to begin with a standard example of asymptotic series.  Consider
\begin{equation}
f(x) = \int_0^{\infty} \frac {e^{-t}}{x + t} dt,
\label{fx1}
\end{equation}
for $x \rightarrow \infty$.  The ``normal'' procedure is to expand the denominator
and write formally
\begin{eqnarray}
f(x) &=& \int_0^{\infty} \left ( {1 \over x} - {t \over x^2} + {t^2 \over x^3} - ...
\right ) \, e^{-t} \, dt\nonumber\\
     &=& {1 \over x} - {1 \over x^2} + ... + (-1)^{n-1} \frac {(n-1)!}{x^n} + ...
\label{fx2}
\end{eqnarray}
The right hand side is clearly a divergent series for any finite value of x.  This
is not surprising because the expansion of the denominator is allowed only for
$t < x$, but the integration over $t$ extends to $\infty$.

Nevertheless one can distill from Eq.~(\ref{fx1}) a different type of expansion which 
avoids the rigid rules of the above power series approach.  Here we find the 
asymptotic series by repeatedly integrating by parts:
\begin{equation}
f(x) = \left. -{e^{-t} \over x+t} \right|_0 ^{\infty} - \int_0^{\infty} \frac {e^{-t}}
{(x+t)^2} dt,
\label{fx3}
\end{equation}
giving
\begin{equation}
f(x) = {1 \over x} - {1 \over x^2} + ... + (-1)^{n-1} {(n-1)! \over x^n} 
+ (-1)^n n ! \, \int_0^{\infty} \frac{e^{-t}}{(x+t)^{n+1}} \, dt.
\label{fx4}
\end{equation}
This equation is exact and is valid for any finite $n$.  If we stop at the 
$n$th term, the remnant is given by
\begin{equation}
(-1)^n n ! \int_0^{\infty} \frac{e^{-t}}{(x+t)^{n+1}} dt = \mathcal{O} \left ( 
\frac{n !}{x^{n+1}} \right ).
\label{fx5}
\end{equation}
We get a satisfactory approximation for a sufficiently large $x$ ($x = n$, e.g., 
would do).  This is an asymptotic series, not a convergent series.  If we stop at
the $n$th term, we will have subtracted from the series the infinite part which, 
formally, is given by $\Sigma_n ^{\infty} (-1)^j j \! x^{-(j+1)}$.
Note that in the series expansion the coefficients grow beyond bound
for $n \rightarrow \infty$.  This is overcome by going to a larger $x$, a freedom
not allowed for the "normal" (convergent) power series expansion in $x$.  (The
phenomenon of the coefficients growing with $n$ happens often for asymptotic series,
but it is not a definition of such a series.)

Another example of asymptotic series is provided by the Stieltjes series\cite{bender} 
for the Stieltjes integral
\begin{eqnarray}
y(x) &=& \int_0^{\infty} \frac{e^{-t}}{1 + xt} dt\nonumber\\
     &=&1 - x + 2 ! x^2 + ... + (-1)^n n! x^n + \epsilon_n (x),
\label{stieltjes1}
\end{eqnarray}
where
\begin{equation}
\epsilon_n (x) = (-1)^{n+1} (n+1)! x^{n+1} \int_0^{\infty} (1+xt)^{-n -2} 
e^{-t} dt,
\label{stieltjes2}
\end{equation}
and 
\begin{equation}
|\epsilon_n(x)| \leq (n+1)! x^{n+1} << x^n, \,\,\,\, x \rightarrow 0_{+}.
\label{stieltjes3}
\end{equation} 

Next we move on to divergent series and their finite parts (i.e., Hadamard's "Parti 
Fini").  Here we encounter a similar situation like that treated above for asymptotic
series.  For this discussion, we use examples that will be relevant to the one-loop
field theory calculations in the next section.  Let us consider
\begin{equation}
F' = \int_0^{\infty} \frac {x}{1+x} dx.
\label{Fprime1}
\end{equation}
Partitioning the integration into two parts, we can write
\begin{equation}
F' = \int_0 ^a \frac {x}{1+x} dx + \int_a ^{\infty} \frac {x}{1+x} dx,
\label{Fprime2}
\end{equation}
for any finite $a$.  The second term contains the infinite part and the approximation
of $F'$ by the finite part is improved by increasing $a$.  But we cannot take $a =
\infty$.  We notice that here $a$ corresponds to the $n$ in Eq.~(\ref{fx4}).  Doing 
the integration for the finite part, we obtain
\begin{equation}
F' = a - \log (1+a) + \int_a ^{\infty} \frac {x}{1+x} dx,
\label{Fprime3}
\end{equation}
valid for all finite values of $a$.  Increasing $a$ indefinitely and subtracting
the infinite part we get
\begin{equation} 
F' = \int_0^{\infty} \frac{x}{1+x} dx \longrightarrow 0.
\label{Fprime4}
\end{equation}

As another example, we consider
\begin{equation}
F = \int_0^{\infty} \frac{x}{(1+x)^2} \, dx,
\label{F1}
\end{equation}
which is also relevant to the physics we discuss in the next section.  Proceeding
as in the calculation of $F'$ we have
\begin{eqnarray}
F &=& \int_0^a {dx \over (1+x)} - \int_0^a {dx \over (1+x)^2} +
\int_a ^{\infty} \frac {x}{(1+x)^2} dx\nonumber\\
  &=& \log a + \frac{1}{1+a} - 1 + \int_a ^{\infty} {x \over (1+x)^2} dx.
\label{F2}
\end{eqnarray}
Subtracting the infinite part with $a \rightarrow \infty$ yields
\begin{equation}
F = \int_0 ^{\infty} {x \over (1+x)^2} dx \longrightarrow -1.
\label{F3}
\end{equation}

For our purposes,
the above two examples illustrate and encapsulate the essence of neutrix 
calculus developed by J.~G.~van~der~Corput 
(partly based on Hadamard's
earlier work) in connection with asymptotic series and divergent integrals.  By
definition, a neutrix is a class of neglgible functions which satisfy the 
following two conditions:\\
(1). A neutrix is an additive group;\\
(2). It does not contain any constant except $0$.\\
Following van der Corput's lead\cite{vdC}, we illustrate the concepts of neutrix 
calculus with the help of various examples.

Example 1. The neutrix consists of functions which go to $0$ as $x \rightarrow 
\infty$.  We call this neutrix $\epsilon (x)$.

Example 2. This concerns the class of polynomials in $x$ with real coefficients.
The neutrix is the subgoup of those polynomials divisible by $x^2 + 1$.
The procedure gives the complex numbers (and this is apparently how Cauchy defined
complex numbers).

Eample 3. The neutrix $H_{\infty}$ consists of functions
\begin{equation}
\nu(\xi) = U(\xi) + \epsilon (\xi),
\label{Hinfty1}
\end{equation}
where $\epsilon(\xi) \rightarrow 0$ as $\xi \rightarrow \infty$ as 
defined in Example 1, and
\begin{equation}
U(\xi) = c_1 \xi + c_2 \xi^2 + ... + c_0 \log \xi + ...
\label{Hinfty2}
\end{equation}  
This neutrix is denoted by $H_{\infty}$. 

Example 4. 
The following example was considered by Hadamard: 
\begin{equation}
C \int_{\xi}^2 x^{s-1} \, dx =
\left\{
\begin{array}{ll}
C s^{-1} 2^s - C s^{-1} \xi^s & \textrm{for}~ s \neq 0 \\
C \log 2 - C \log \xi & \textrm{for}~ s=0
\end{array}
\right.
\label{ex1}
\end{equation}
Here $\xi > 0$, $s$ is a real number, and $C$ is complex.
For $s > 0$, the integral converges.
%even as $\xi \rightarrow 0$.  
For $s < 0$, $s^{-1} \xi^s$ grows without bound as 
$\xi \rightarrow 0$, and so does $- \log \xi$ for $s=0$.
Hadamard called $C s^{-1} 2^s$ and $C \log 2$ the finite part 
("parti fini"), and 
$C s^{-1} \xi^s$ and $C \log \xi$ the infinite part ("parti infini").
He neglected the infinite part.  For $\xi \rightarrow 0$,
van der Corput wrote the integral as
%For the case of integral
%$s$,
%this neutrix can be
%extended to include functions of the form
%\begin{equation}
%f(\xi) = c_{-N} \xi^{-N} + ... + c_{-1} \xi^{-1} + c_0 \log \xi.
%\label{f(x)}
%\end{equation}
%Then the neutrix, which we call $N(0)$, consists of the functions
%\begin{equation}
%\nu(\xi) = f(\xi) + \epsilon (\xi).
%\label{N(0)}
%\end{equation}
\begin{equation}
C \int_{H_0}^2 x^{s-1} \, dx =
\left\{
\begin{array}{ll}
C s^{-1} 2^s & \textrm{for}~ s \neq 0\\
C \log 2 & \textrm{for}~ s = 0.
\end{array}
\right.
\label{ex11}
\end{equation}
Note the analytic extension in the complex $s$ plane of the answer
for Re $s > 0$ to the entire complex plane with the exclusion of
$s=0$.

In terms of the neutrix $H_{\infty}$ defined in Example 3, we can rewrite
Eqs.~(\ref{Fprime4}) and (\ref{F3}) as
\begin{equation}
\int_0^{H_{\infty}} \frac{x}{1+x} dx = 0,
\label{Fprime5}
\end{equation}
and
\begin{equation}
\int_0^{H_{\infty}} \frac{x}{(1+x)^2} dx = -1,
\label{F4}
\end{equation}
respectively.

We conclude this section by pointing out that neutrices are essential
to find the coefficients of an asymptotic series of the type
\begin{equation}
f(x) = ... + c_2 x^2 + c_1 x + c_0 + c_{0l} \log x + ... 
+ c_{-1} x^{-1} + c_{-2} x^{-2} + ...
+ c_{-1l} x^{-1} \log x + ...
\label{coeff1}
\end{equation}
To find the coefficient $c_{-1}$, for example, we multiply both sides by
$x$ and take the Hadamard $H_0$ limit at $x=0$.  On the right hand side of
Eq.~(\ref{coeff1}), the terms to the left of $c_{-1}$ are part of 
$\epsilon (0)$ and hence converge to zero.  On the other hand, the terms
to the right of $c_{-1}$ are all in $H_0$ and hence are neglected.  We
conclude
\begin{equation}
c_{-1} = \lim_{H_0} x f(x).
\label{coeff2}
\end{equation}

More mathematical properties of neutrices and asymptotic series can be found
in the two appendices.

\bigskip

\section{\bf Neutrix calculus applied to 1-loop QED and $\phi^4$ theory}

As the first example in the application of neutrices to quantum field theory,
let us consider QED:
\begin{equation}
\mathcal{L} = - \bar{\psi} [\gamma \cdot ( {1 \over i}\partial
- eA) + m] \psi 
- {1 \over 4} (\partial_{\mu} A_{\nu} -  \partial_{\nu} A_{\mu})^2.
%(\partial^{mu} A^{\nu} - \partial^{nu} A^{mu}).
\label{QEDL}
\end{equation}
We begin with
the one-loop contribution to the electron's self energy in the Feynman gauge,
\begin{equation}
\Sigma (p) = -ie^2 \int {d^4k \over (2 \pi)^4} \frac {\gamma_{\mu}
 [- \gamma \cdot (p-k)+m] \gamma^{\mu}}{[k^2 + \lambda^2][(p-k)^2 + m^2]},
\label{sigma}
\end{equation}
where $m$ is the electron bare mass and we have given the photon a fictitious
mass $\lambda$ to regularize infrared divergences. Expanding $\Sigma(p)$
about $\gamma \cdot p = -m$,
\begin{equation}
\Sigma (p) = \mathcal{A} + \mathcal{B} (\gamma \cdot p + m) + \mathcal{R},
\label{sigma1}
\end{equation}
one finds (cf. results found in Ref.\cite{JR})
\begin{equation}
\mathcal{A} = -{ \alpha \over 2 \pi} m \left({3 \over 2} D + {9 \over 4}
\right),
\label{sigma2}
\end{equation}
\begin{equation}
\mathcal{B} = - {\alpha \over 4 \pi} \left( D - 4 \int_{\lambda \over m}
^1 {dx \over x} + {11 \over 2} \right),
\label{sigma3}
\end{equation}
where $\alpha = e^2/4 \pi$ is the fine structure constant and
$D$ is given by 
\begin{eqnarray}
D &=& {1 \over i \pi^2} \int {d^4k \over (k^2 + m^2 )^2}\nonumber\\
&=& \int_0^\infty {k^2 dk^2 \over (k^2 +m^2)^2},
\label{sigma4}
\end{eqnarray}
with the second expression of $D$
obtained after a Wick rotation.
%Eq.~(\ref{sigma4}) and Eq.~(\ref{D})
%in the pre-neutralized and neutralized forms respectively.
%\begin{eqnarray}
%D &=& {1 \over i \pi^2} \int {d^4k \over (k^2 + m^2 )^2}\nonumber\\
%&=& \int_0^\infty {k^2 dk^2 \over (k^2 +m^2)^2}.
%\label{sigma4}
%\end{eqnarray}
We note that $\mathcal{R}$, the last piece of $\Sigma (p)$ in
Eq.~(\ref{sigma1}), is finite.
%and the second expression of $D$ in
%Eq.~(\ref{sigma4}) is obtained after a Wick rotation.
Mass renormalization and wavefunction renormalization are given by
$m_{ren} = m - \mathcal{A}$ and $\psi_{ren} = Z_2^{-1/2} \psi$ respectively
with $Z_2^{-1} = 1 - \mathcal{B}$.
%It is in the calculation of the logarithmically divergent $D$
%where we apply neutrix calculus.
%Introducing dimensionless variable $q = k^2/m^2$, we
%bring in $H_\infty$ to write $D$ as
%\begin{equation}
%D = \int_0^{H_{\infty}} {q dq \over (q + 1)^2} = -1,
%\label{D}
%\end{equation}
%where we have recalled that, for $q \rightarrow \infty$, $\log q$ is
%negligible in the Hadamard neutrix $H_\infty$.
Now, 
introducing dimensionless variable $q = k^2/m^2$, we
bring in the neutrix $H_\infty$ to write $D$ as
\begin{equation}
D = \int_0^{H_{\infty}} {q dq \over (q + 1)^2} = -1,
\label{D}
\end{equation}
where, for the last step, we have used Eq.~(\ref{F4}). 
%recalled that, for $q \rightarrow \infty$, $\log q$ is
%negligible in the Hadamard neutrix $H_\infty$.  
Since $D = -1$ is finite,
it is abundantly clear that the renormalizations are {\it finite} in the
framework of neutrix calculus.  There is no need for
a separate discussion of the electron vertex function renormalization
constant $Z_1$ due to the Ward identity $Z_1 =Z_2$.

The one-loop contribution to vacuum polarization is given by
\begin{equation}
\Pi_{\mu \nu} (k) = i e^2 \int {d^4p \over (2 \pi)^4} Tr \left (
\gamma_{\mu} {1 \over \gamma \cdot (p + \frac{k}{2}) + m}
\gamma_{\nu} {1 \over \gamma \cdot (p - \frac{k}{2}) + m} \right ).
\label{pi}
\end{equation}
A standard calculation\cite{JR} shows that $\Pi_{\mu \nu}$ takes on the form
\begin{equation}
\Pi_{\mu \nu} = \delta m^2 \eta_{\mu \nu} + (k^2 \eta_{\mu \nu} - k_{\mu}
k_{\nu}) \Pi(k^2),
\label{pi1}
\end{equation}
where $\eta_{\mu \nu}$ is the flat metric (+++-),
\begin{equation}
\delta m^2 = {\alpha \over 2 \pi} (m^2 D + D'),
\label{pi2}
\end{equation}
and
\begin{equation}
\Pi(k^2) = - {\alpha \over 3 \pi} (D + \frac{5}{6}) +
{2 \alpha \over \pi} \int_0^1 dx \, x (1-x) \,
\log \left (1 + {k^2 \over m^2} x(1-x) \right ),
\label{pi3}
\end{equation}
with
\begin{equation}
D' = {1 \over i \pi^2} \int \frac {d^4p}{p^2 + m^2},
\label{pi4}
\end{equation}
and $D$ given by Eq.~(\ref{sigma4}).  Just as $D$ is rendered finite upon
invoking neutrix calculus (see Eq.~(\ref{D})), so is $D'$:
\begin{equation}
D' = m^2 \int_0^{H_{\infty}} {q dq \over q + 1} = 0,
\label{pi5}
\end{equation}
%since both $q$ and $\log q$, for $q \rightarrow \infty$, are
%negligible in $H_\infty$.
where, for the last step, we have used Eq.~(\ref{Fprime5}).
Thus neutrix calculus yields a finite renormalization for both the photon
mass and the photon wavefunction $A_{ren}^{\mu} = Z_3 ^{-1/2} A^{\mu}$
(and consequently also for charge $e_{ren} = Z_3^{1/2} e$) where $Z_3^{-1} = 1
- \Pi(0)$.
%To further show that the neutrix calculus yields the correct physics, let us
%rederive the running of the coupling with energy-momentum\cite{run}
%in the neutrix framework.  Consider
In electron-electron scattering by the
exchange of a photon with energy-momentum $k$, vacuum polarization effects
effectively replace $e^2$ by $e^2/(1 - \Pi(k^2))$, i.e.,
\begin{eqnarray}
e^2 \rightarrow e_{eff}^2 &=& \frac {e^2}{1 - \Pi (k^2)}\nonumber\\
&=& \frac {e_{ren}^2}{Z_3 (1 - \Pi (k^2))}\nonumber\\
&=& \frac {e_{ren}^2} {1 - (\Pi(k^2) - \Pi(0))}.
\label{run1}
\end{eqnarray}
Eq.~(\ref{pi3}) can be used, for $k^2 \gg m^2$, to show that
\begin{equation}
\alpha_{eff} (k^2) = \frac {\alpha}{1 - {\alpha \over 3 \pi} \log \left (
{k^2 \over \exp(5/3)\, m^2} \right )}.
\label{run2}
\end{equation}
Thus we have obtained the correct running of the coupling\cite{PS}
with energy-momentum in
the framework of neutrices.  In fact, the {\it only} effect of neutrix calculus,
when applied to QED (and other renormalizable theories), is to convert
{\it infinite} renormalizations (obtained without using neutrix calculus) to
mathematically well-defined {\it finite} renormalizations.
As far as we can tell, {\it all} (finite)
physically observable results of QED are recovered.
In passing we mention that the use of
neutralized integrals does not affect the results of axial triangle anomalies.

Let us now apply neutrix calculus to the $\phi^4$ theory,
\begin{equation}
\mathcal{L} = - {1 \over 2} \partial^{\mu} \phi \, \partial_{\mu} \phi
- {1 \over 2} m^2 \phi^2 - {\lambda \over 4 !} \phi ^4.
\label{phi4L}
\end{equation} 
The one-loop self-energy is given by
\begin{equation}
\Sigma (p) = i {\lambda \over 2} \int {d^4k \over (2 \pi)^4}
\frac{1}{k^2 + m^2},
\label{phi1}
\end{equation}
which, by Eqs.~(\ref{pi4}) and (\ref{pi5}), vanishes!  Thus there is no
mass renormalization ($m_{ren}^2  = m^2 - \Sigma (0) = m^2$) and no 
wave-function renormalization (wave-function renormalization constant 
$Z_{\phi} = 1$) in the one-loop approximation.  The four-point function
$\Gamma (s, t, u)$, for incoming particle momenta $p_i$, receives
the following  one-loop contributions
\begin{equation}
\Gamma (s,t,u) = {\lambda^2 \over 2} \int \frac{d^4 p}{(2 \pi)^4}
{1 \over p^2 + m^2} {1 \over (p-q)^2 + m^2} + \textrm{(2 crossed terms)}.
\label{phi2}
\end{equation}
Here $q = p_1 + p_2$, and we have used the Mandelstam variables $s = -q^2$, 
$t = -(p_1 + P_3)^2$, and $u = -(p_1 + p_4)^2$.  A straight-forward calculation
yields
\begin{equation}
\Gamma (s,t,u) = i \frac{ \lambda^2}{32 \pi^2} \left\{ 3D 
- \int_0^1 dz \left( \log( 1 - {s \over m^2} z[1-z]) + (s \rightarrow t)
+ (s \rightarrow u) \right) \right\}.
\label{phi3}
\end{equation}
Thus, the coupling receives a finite renormalization
\begin{eqnarray}
\lambda_{ren} &=& \lambda \left( 1 - {3D \lambda \over 32 \pi^2} \right )
\nonumber\\
              &=& \lambda \left ( 1 + {3 \lambda \over 32 \pi^2} \right ),
\label{phi4}
\end{eqnarray}
where we have defined $\lambda_{ren}$ by $\Gamma(s,t,u)$ at $s=t=u=0$, and have
used $D =-1$, given by Eq.~(\ref{D}), for the last step.

\bigskip

\section{\bf Dimensional regularization and pure Yang-Mills theory}

As shown by the appearance of photon mass in the above discussion of
vacuum polarization in QED, the application of neutrix calculus to the
energy-momentum cutoff regularization scheme is 
not too convenient to use for more complicated
theories like those involving Yang-Mills fields.  For those theories, one
should use other regularization schemes that manifestly preserve the 
Ward identity.
% like the dimensional regularization\cite{PS}.
% developed by 't Hooft and Veltman.
In this regard, we note that already in 1961, van der Corput suggested that,
instead of finding the appropriate neutrices, one can continue analytically
in any variable (presumably including the dimension of integrations)
contained in the problem of tackling apparent divergences to calculate
the coefficients of the corresponding asymptotic series.  It so happened 
that this was the approach taken by 't Hooft and Veltman who spearheaded
the use of dimensional regularizations\cite{PS}.   Let us now explore using
neutrix calculus in conjunction with the dimensional regularization scheme.
In that case, negligible functions will include $1/\epsilon$ where
$\epsilon = 4 - n$ is the deviation of spacetime dimensions from 4.
In the calculations, the
internal energy-momentum integration is now over n dimensions.  
The one-loop contributions to the electron's self-energy in QED is given by
\begin{equation}
\Sigma (p) = -\frac{e^2}{(4\pi)^{n/2}} \Gamma(2 - {n \over 2})
\int_0^1 dx \frac {(n-2) \gamma \cdot p (1-x) + nm}
{[p^2 x(1-x) + m^2x + \lambda^2 (1-x)]^{2 - n/2}}.
\label{newsigma}
\end{equation}
Again we expand $\Sigma(p)$ about $\gamma \cdot p = -m$ as in 
Eq.~(\ref{sigma1}).
Using the approximation for the gamma function
\begin{equation}
\lim_{\epsilon \rightarrow 0} \Gamma(\epsilon) =
{1 \over \epsilon} - \gamma + \mathcal{O} (\epsilon),
\label{dim}
\end{equation}
where $\gamma \simeq .577$ is the Euler-Mascheroni constant, 
and the approximation 
\begin{equation}
f^{\epsilon} \simeq 1 + \epsilon \log f,
\label{fepsilon}
\end{equation}
for $\epsilon \ll 1$,
one finds
\begin{eqnarray}
\mathcal{A} &=& \frac {\alpha m} {4 \pi} [3(\gamma - \log 4 \pi) +1]
+{\alpha \over 2 \pi} m \int_0^1 dx (1+x) \log D_0,\nonumber\\
%\end{equation}
%\begin{equation}
\mathcal{B} &=& {\alpha \over 4 \pi}[1 + \gamma - \log 4 \pi] 
+ {\alpha m^2 \over \pi}
\int_0^1 dx \frac{x(1 - x^2)}{m^2 x^2 + \lambda^2 (1-x)}
+ {\alpha \over 2 \pi}\int_0^1 dx (1-x) \log D_0,
\label{dim0}
\end{eqnarray}
where $D_0 = m^2x^2 + \lambda^2(1-x)$, \footnote{Since $D_0$ is 
dimensionful, the appearance of $\log D_0$ is very strange.  This is
due to the fact that the $e^2$ in Eq.~(\ref{newsigma}) is in fact
also dimensionful for $\epsilon \neq 0$.  
If we replace $e^2$ there by $e^2 \mu^{4 - n}$ with
a certain mass $\mu$ so that the $e^2$ will then be dimensionless, $D_0$
will be replaced by dimensionless $D_0/\mu^2$.  For simplicity we have 
not written this dimensional dependence out explicitly.  Hereafter this 
dependence will be left implicit.}
and we have invoked neutrix
calculus in dropping $-3 \alpha m/(2 \pi \epsilon)$ and $- \alpha /
(2 \pi \epsilon)$ from $\mathcal{A}$ and $\mathcal{B}$ respectively.
(In passing we note that $\mathcal{R}$ in the expansion of $\Sigma(p)$ in 
the dimensional regularization scheme is the same as in the 
energy-momentum regularization used in the last section.)

The one-loop vacuum polarization takes on the same form as given
by Eq.~(\ref{pi1}), but now with $\delta m^2 = 0$, and
\begin{equation}
\Pi(k^2) = -8 e^2 {\Gamma(2- {n \over 2}) \over (4\pi)^{n/2}}
\int_0^1dx \frac {x(1-x)}{[k^2 x(1-x) + m^2]^{2 -n/2}}.
\label{dim03}
\end{equation}
In the limit $\epsilon = 4 - n \rightarrow 0$ we obtain
\begin{eqnarray}
\Pi (k^2)      &=& {\alpha \over 3 \pi} [\gamma - \log 4 \pi] 
+ {2 \alpha \over \pi}
\int_0^1 dx \, x(1-x)\, \log [m^2 + x(1-x)k^2],\nonumber\\
%\end{equation}
%\begin{equation}
\frac {1}{Z_3} &=& 1 - {\alpha \over 3 \pi}[\gamma - \log 4 \pi]
- {\alpha \over 3 \pi} \log m^2,
\label{dim1}
\end{eqnarray}
where we have followed neutrix calculus in dropping $- e^2/(6 \pi^2 
\epsilon)$ from $\Pi(k^2)$. 
By design, the generalized neutrix calculus renders all the renormalizations
{\it finite}.  Again, {\it all} physically measurable results of QED appear
to be recovered.

Next we consider pure non-Abelian gauge theroy for a group $G$ with structure
constants $C_{abc}$ 
\begin{equation}
\mathcal{L} = -{1 \over 4} (\partial^{\mu} A_a^{\nu} - \partial^{\nu} A_a^{\mu}
+g_0 C_{abc} A_b^{\mu} A_c^{\nu})^2.
\label{nAg}
\end{equation}
The one-loop contribution to the self-energy of the gauge field, in the Feynman 
gauge,
is given by
\begin{equation}
\Pi_{\mu \nu}^{cd} (k) = (k^2 \eta_{\mu \nu} - k_{\mu} k_{\nu}) \Pi^{cd} (k),
\label{Pi1}
\end{equation}
with
\begin{equation}
\Pi^{cd} (k) = C_{cd}^2 \frac{g_0^2}{2(4 \pi)^{n/2}} 
\Gamma(2 - {n \over 2}) \int_0^1 dx \, \mathcal{D}^{{n \over 2} - 2}
[(4 - 2n) + (6n -4)x + (8 - 4n) x^2],
\label{Pi2}
\end{equation}
where the symbol $C_{cd}^2$ stands for $C_{abc}C_{abd}$ and
$\mathcal{D} = k^2 x(1-x)$.

With the aid of 
Eqs.~(\ref{dim}) and (\ref{fepsilon}), we get 
%$f^{\epsilon} \simeq 1 + \epsilon \log f$ in the limit
%$\epsilon \rightarrow 0$, $\Pi^{cd} (k)$ can be shown to be
\begin{equation}
\Pi^{cd} (k) = \frac{C_{cd}^2 g_0^2}{32 \pi^2} \left( {20 \over 3} {1 \over 
\epsilon} - {10 \over 3} \log k^2 + {10 \over 3}[-\gamma + \log 4 \pi]
+ 7 - {1 \over 9} \right ).
\label{Pi3}
\end{equation}
Invoking neutrix calculus to remove the negligible $\epsilon^{-1}$ term and
specializing to the $SU(N)$ group where $C_{cd}^2 = N \delta_{cd}$, we obtain
\begin{equation} 
\Pi^{cd}(k) = \delta_{cd} \Pi(k^2),
\label{Pi4}
\end{equation}
with
\begin{equation}
\Pi(k^2) \rightarrow \frac{N g_0^2}{32 \pi^2} \left( - {10 \over 3} \log k^2 
+ ... \right),
\label{Pi5}
\end{equation}
where $ ...$ stands for the finite $k$-indepentdent terms in Eq.~(\ref{Pi3}).
For QED we have used the on-shell renormalizations with massive electrons.  
For pure Yang-Mills theory we will do the renormalizations for massless
gauge fields at space-like energy-momentum $\mu^2$.  The renormalization
constant $Z_3$ is then given by
\begin{equation}
Z_3 = \frac {1}{1 - \Pi(\mu^2)} \simeq 1 - {10 \over 3} {N g_0^2 \over 
16 \pi^2} \log \mu + ...
\label{Pi6}
\end{equation}
Similarly the renormalization constant for the three-point function in the
Feynman gauge can be calculated to yield
\begin{equation}
Z_1 \simeq 1 - {4 \over 3} {N g_0^2 \over 16 \pi^2} \log \mu + ...
\label{Pi7}
\end{equation}
where $...$ stands for finite $\mu$-independent terms.  
The Callan-Symanzik function for the renormalized coupling 
$g = Z_1^{-1} Z_3^{3/2} g_0$
in the one-loop approximation is then given by
\begin{eqnarray}
\beta (g) &=& \mu {\partial g \over \partial \mu}\nonumber\\
          &=& - {11 \over 3} {N g^3 \over 16 \pi^2}.
\label{betafn}
\end{eqnarray}
Thus we have recovered the negative $\beta$ function for $SU(N)$ Yang-Mills 
theory.

\bigskip

\section{\bf Conclusions and discussions}

In this paper, we have proposed to apply neutrix calculus,
in conjunction with Hadamard integrals,
developed by J.G. van der Corput \cite{vdC}
in connection with asymptotic
series, to quantum field theory,
to obtain finite results for the coefficients in perturbation series.
The replacement of regular integrals by Hadamard integrals in
quantum field theory appears to make good mathematical sense, as
van der Corput observed that Hadamard integrals are the proper tool to
calculate the coefficients of an asymptotic series.
(Actually Hadamard integrals work equally well for convergent series.)

For renormalizable quantum field theories like QED, $\phi^4$ theory, and 
pure Yang-Mills, we have demonstrated (even if not adequately) 
that we recover all the usual physically observable results.  
The only effect of neutrix calculus appears to just change the ``amount'' of
renormalizations ---from infinite renormalizatins to finite
renormalizations.  But what about non-renormalizable field theories like
quantum gravity?
Using dimensional regularization, 'tHooft and Veltman\cite{toast} found that
pure gravity is one-loop renormalizable, but in the presence of a scalar
field, renormalization was lost.  For the latter case, they found that the
counterterm evaluated on the mass shell is given by
$\sim \epsilon^{-1} \sqrt{g} R^2$ with $R$ being the Ricci scalar.  Similar
results for the cases of Maxwell fields and Dirac fields etc
(supplementing the Einstien field) were obtained\cite{deser}.  It is
natural to inquire whether the application of neutrix calculus could improve
the situation.  Our preliminary result is that now essentially the
divergent $\epsilon^{-1}$ factor is replaced by $-\gamma +$ finite terms.
%It is not clear to us at this point whether we recover this finite result
%if we use another regularization scheme other than the dimensional
%regularization.

Since neutrix calculus does not tolerate infinities, we conjecture that it
can also be used to ameliorate other problems that can be 
traced to ultraviolet divergences.  Two well-known problems come to mind.
The hierarchy problem in particle physics is due to the fact that
the Higgs scalar self-energies diverge quadratically.  It is conceivable
that neutrix calculus may help.
% leading to a
%stability problem in the standard model of particle physics.  But
%neutrix calculus treats quadratic divergences no different from
%logarithmic
%divergences, since both divergences belong to (the negligible functions of)
%the neutrix.
%Neutrix calculus may also ameliorate the cosmological constant problem in
%quantum gravity.
The cosmological constant problem can be traced to the quartic
divergences in zero-point fluctuations from all quantum fields.  
%But again,
%neutrix calculus
%treats quartic divergences no different from logarithmic divergences.
Perhaps neutrix calculus can help with this problem too.
Indeed, for a theory of gravitation with a cosmological constant term,
the cosmological constant receives at most a finite renormalization from
the quantum loops in the framework of neutrix calculus.

%We conclude with a comment on what neutrix calculus means to the general
%question of renormalizability of a theory.  We recall that a theory is
%renormalizable if, in loop calculations, the counterterms vanish or if they
%are proportional to terms in the original Lagrangian (the usual
%renormalization through rescaling).  It is still renormalizable if, to all
%loops, the counterterms are of a new form, but only a finite number of such
%terms exist.  By this standard, neutrix calculus does not change the
%renormalizability of a theory, since it merely changes potentially infinite
%renormalizations to finite renormalizations.  On the other hand,
%non-renormalizable terms, i.e., terms with positive superficial degree
%of divergence, are tolerated in neutralized quantum field theory.
By ridding quantum field theories of their ultraviolet divergences, neutrix
calculus may appear to have fallen victim to its own successes, for now we have 
lost renormalizability as a physical restrictive criterion in the choice of
sensible theories.  However, we believe that this is actually not as
big a loss as it may first appear.  Quite likely, all realistic theories now
in our possession are actually effective field
theories.\cite{Schwinger,Weinberg}  They appear to be
renormalizable field theories because, at energies now accessible, or
more correctly, at sufficiently low energies, all the non-renormalizable
interactions are highly suppressed.  By tolerating
non-renormalizable theoriess, neutrix calculus has provided us with a more 
flexible framework to study all kinds of particle interactions. 
%freed us from the past dogmatic and rigid requirement
%of renormalizability.  (Having said that, given a choice between
%renormalizable field theories and effective field theories, we
%still prefer the
%former to the latter because of the former's compactness and predictive power.
%But the point is that both types of theories can be accommodated in the
%framework of neutrix calculus.)  Furthermore, if the
%application of neutrix calculus to loop calculations results in a
%term of a new form (like the Pauli term in QED) that is finite, then
%we have a prediction which, in principle, can be checked
%against experiments to confirm or invalidate the theory in question.  For the
%latter case, we will have to modify the theory by including a term of
%that form in the Lagrangian,
%making the parameter associated with the new term an adjustable parameter
%rather than one that is predicted by the theory.  This loss of predictive
%power is again not as big a loss as one may dread.
%Lastly we should emphasize
%that, for renormalizable theories as well as
%non-renormalizable theories (like quantum gravity?), neutrix calculus is
%a useful tool 

To the extent that it is relevant for asymptotic series
and lessens the divergences in quantum field theories, neutrix calculus 
is a powerful 
tool that is, in our opinion, too valuable not to make use of.
% With the
%help of neutrix calculus, after we say, ``The perturbative
%expansion in question is meaningful because the loop corrections are
%progressively small,'' we no longer have to
%add sheepishly, tongue in cheek, ``though (apparently) infinite.''
How much it can really help to ameliorate problems related to ultraviolet 
divergences in quantum field theory remains to be seen.  But based on our
study so far, we tentatively argue that 
neutrix calculus has banished infinities from quantum field theory.

\bigskip

\section*{Acknowledgments}

We thank J.J. Duistermaat, E. M. de Jager, T. Levelt, and
T. W. Ruijgrok for encouragement and for kindly providing us with relevant
references
of the work by J.G. van der Corput.  We thank C. Bender, K. A. Milton,
and J. Stasheff for
useful
discussions.
We also thank L.~Ng and X.~Calmet for
their help in the preparation of this manuscript.  We are especially 
indebted to Prof. E.~M.~de~Jager for suggesting to us 
his way to introduce the subject
of neutrix calculus.
We are grateful to the late Paul Dirac
and Julian Schwinger for inspiring us to look for a better mathematical
foundation for quantum field theory.
This work was supported in part by DOE
and by the Bahnson Fund of University of North
Carolina.

\bigskip

\begin{center}
{\bf Appendix A: The Hadamard neutrix $H_a$ and Hadamard series expansion}
\end{center}

In Section II, we considered some examples of asymtotic series.  Essentially,
the series $f(x) = a_0 + a_1 (x - b) + a_2 (x-b)^2 + ...$ for finite $b$ is an
asymptotic series \cite{bender}
if and only if there exists an $n_0 > 0$, such that
for $n > n_0$,
%\begin{equation}
%\[ \lim_{x \rightarrow b}  \frac {1}{(x-b)^n} \left | f(x) - a_0 -a_1(x-b)
%- ... - a_n (x-b)^n \right | = 0 \],\nonumber
%\label{asym}
%\end{equation}
\begin{equation}
\lim_{x \rightarrow b} \frac{1}{(x-b)^n} \left| f(x) - a_0 - a_1(x-b)
- \dots - a_n (x-b)^n \right| = 0,
\label{asym}
\end{equation}
with the remnant being at most $\sim (x-b)^{n+1}$.

In this appendix, we consider the Hadamard series expansion and a generalized 
definition of asymptotic series.  Let $\kappa$ be a region in the complex 
plane with a finite limit point $a$ which does not belong to $\kappa$.  A 
crucial property is that a point $\xi$ in $\kappa$ can approach $a$ in such 
a way that arg$(\xi - a)$ has a finite limit.  We say that $U(\xi)$ in $\kappa$
has a Hadamard expansion in powers of $\xi - a$ if $U(\xi)$ is defined in
$\kappa$, and for points $\xi$ near $a$, has an asymptotic expansion of the 
kind
\begin{equation}
U(\xi) \sim \Sigma_{h=0}^{\infty} \chi_h \, (\xi - a)^{\Psi_h} \,
\log^{k_h} (\xi-a).
\label{U(x)}
\end{equation}
Here $\chi_h$, $\Psi_h$ and integers $k_h \geq 0$ are independent
of $\xi$, Re $\Psi_h \rightarrow \infty$ as $h \rightarrow
\infty$, and $\log^{k_h} (\xi)$ stands for $(\log (\xi))^{k_h}$.
%HOW TO PUT ACCENT IN SIGMA?

The definition of asymptotic series here is more powerful than the one 
we used in Section~II in that, for every number $q$ (independent of $\xi$),
there exists an integer $m_0 \geq 0$ with the property that, for every 
number $m \geq m_0$, there are two positive numbers $c_m$ and $\epsilon_m$
such that, for every point $\xi$ in $\kappa$ with $|\xi - a| < \epsilon_m$,
\begin{equation}
\left |U(\xi) - \Sigma_{h=0}^{m-1} \chi_h \, (\xi - a)^{\Psi_h} \,
\log^{k_h} (\xi-a) \right | < c_m |\xi - a |^q.
\label{asII}
\end{equation}
The new definition of asymptotic series is essential in deriving the
analytic properties of functions represented by neutralized integrals.
Note that this means that there is a radius of convergence which goes to
zero as $m$ goes to infinity.  If, in Eq.~(\ref{asII}), there is no
term with $\Psi_h = k_h = 0$, then we speak of a Hadamard series expansion
without constant term.

We now define the Hadamard neutrix $H_a$.  It consists of all functions
$\nu(\xi)$, which in the neighborhood of $a$ can be written as
\begin{equation}
\nu(\xi) = U (\xi) + \epsilon (\xi).
\label{hadneut1}
\end{equation}
Here $U(\xi)$ has a Hadamard series expansion in powers of $(\xi -a)$ without
a constant term and $\epsilon(\xi)$ approaches zero when $\xi$ in $\kappa$ 
approaches $a$.  This neutrix is called $H_a$; it has a domain $\kappa$, a 
variable $\xi$, and $a$ is called the carrier.  Thus every negligible functions
in $H_a$ can be written as
\begin{equation}
\Sigma_{h=0}^{\infty} \chi_h \, (\xi - a)^{\Psi_h} \,
\log^{k_h} (\xi-a) + \epsilon(\xi),
\label{Haform}
\end{equation}
with no $\Psi_h = k_h = 0$ term.

We end Appendix A with three examples.

Example 1.  If two different points $a$ and $b$ are connected by a rectifiable
curve in the complex plane, which has a tangent in $a$ but does not contain
$a$, then for every integer $k \geq 0$ and for every complex $s$
\begin{equation}
\int_{H_a}^b (z - a)^{s-1} \log^k(z-a) \, dz =
\left \{
\begin{array}{ll} 
{1 \over k+1} \log^{k+1} (b-a) & \textrm{for}~ s=0\\
\left ({\partial\over \partial s} \right)^k {(b-a)^s \over s} 
& \textrm{for}~ s \neq 0
\end{array}
\right.
\label{haex1}
\end{equation}

Example 2.  For every complex s, except $0, -1, -2, ...$,
\begin{equation}
\int_{H_0}^{\infty} x^{s-1} \, e^{-x} \, dx = \Gamma(s).
\label{haex2}
\end{equation}

Example 3.  For $-\pi <$ arg $l < \pi$, $s$ not an integer $\leq 0$, and 
$s+t$ not an integer $\geq 0$, one has
\begin{equation}
\int_{H_0}^{H_{\infty}} x^{\lambda s -1} \, ( l + x^{\lambda})^t \,
\log^kx \, dx 
= {1 \over \lambda^{k+1}} \left( {\partial \over \partial s} \right )^k
\frac{\Gamma(s) \Gamma(-s-t)}{\Gamma(-t)} l^{s + t}.
\label{haex3a}
\end{equation}
Here the integral is along the positive axis.
%and $H_{\infty}$ has 
%$parameter $0$.
This integral contains the special case (with $\lambda =2, k=0, 
t = -\alpha$, and $2s = \beta + 1$) 
\begin{equation}
\int_{H_0}^{H_{\infty}} \frac{x^{\beta}}{(x^2 + l)^{\alpha}} dx
={1 \over 2} \frac{\Gamma \left( {\beta +1 \over 2} \right)
\Gamma \left( \alpha - { \beta + 1 \over 2} \right)}
{\Gamma(\alpha) l^{\alpha -(\beta + 1)/2}},
\label{haex3b}
\end{equation}
an expression encountered in dimensional regularization calculations
for which $\beta = 3 - \epsilon$ with a positive infinitesimally small 
$\epsilon$.

%Before applying neutrices to QED, we need to consider the generalized
%Hadamard neutrix $H_a$ defined to contain the negligible functions
%\begin{equation}
%functions $U(\xi)$ is defined by an asymptotic series based on $a$:
%An example is provided by
%\begin{equation}
%\int_{H_a}^b (z-a)^{-1} \, \log^k (z-a) \, dz = (k+1)^{-1} \log ^{k+1} (b-a).
%\label{exHa}
%\end{equation}
%Similarly, we can define the Hadamard neutrix $H_\infty$ by Eq.~(\ref{H(a)})
%where now $\epsilon(\xi) \rightarrow 0$ as $\xi \rightarrow \infty$ and
%the function $U(\xi)$ has a Hadamard development in powers of $\xi^{-1}$ in
%its asymptotic series:
%\begin{equation}
%U(\xi) \sim \Sigma_{h=0}^{\infty} \chi_h \, \xi^{\Psi_h} \,
%\log^{k_h} \xi,
%\label{U1(x)}
%\end{equation}
%where  Re $\Psi_h \rightarrow -\infty$ as $h \rightarrow
%\infty$.

\bigskip

\begin{center}
{\bf Appendix B: Neutrices and products of singular functions}
\end{center}

It has often been claimed that infinities find their way into quantum field
theory because one works with products of singular functions (like Feynman 
propagators) in the theory.  In what used to be called axiomatic field 
theory, one had to search for the definitions of the products of 
operator-valued distributions.  But whereas generalized functions usually 
cannot be multiplied in the theory of distributions developed by Schwartz,
the Hadamard approach adopted in this paper allows the multiplication 
for a wide class of distributions as we will show in this Appendix.  In fact
this is the reason that one obtains finite rather than infinite 
renormalizations in the framework of neutrix calculus.

Lighthill\cite{Lighthill} has already defined and listed
the products for some distributions in his book.  Van der Corput\cite{vdC} 
has been able to simplify and extend Lighthill's list by doing
what one usually does in quantum field theory, namely, by using Fourier
transformed (FT) functions
\begin{equation}
FT[f(x)] = F(p) = \int_{-\infty}^{\infty} e^{-ipx} f(x) \, dx.
\label{Fourier}
\end{equation}
Then he defines the FT of the product of two functions by folding (as usual),
but using Hadamard integtrals (instead of the ``normal'' integrals):
\begin{equation}
FT[f_1(x) \, f_2(x)] = \int_{H_{-\infty}}^{H_{\infty}} 
{dk  \over 2 \pi} F_1 (k) \, F_2 (p-k).
\label{FT2}
\end{equation}
  
Consider, for example,
the one-dimensional Dirac delta function multiplying itself
$\delta (x) \times \delta (x)$.  In Schwartz' approach,
this product is not mathematically
meaningful because its Fourier transform diverges:
\begin{equation}
\int_{-\infty}^{\infty} {dk \over 2 \pi} \, 1 \times 1 \,
\longrightarrow \infty,
\label{delsq}
\end{equation}
where we have used the convolution rule in Fourier transform
and have noted that the
Fourier transform of the Dirac delta function is $1$.  In contrast,  
the Hadamard-neutralized Fourier transform of the product
$\delta (x) \times \delta (x)$, 
\begin{equation}
\int_{H_{-\infty}}^{H_{\infty}} {dk \over 2 \pi} \, 1 \times 1 \, = 0,
\label{delsq2}
\end{equation}
yields
\begin{equation}
\delta (x) \times \delta (x) = 0,
\label{delsq3}
\end{equation}
a mathematically meaningful (though somewhat counter-intuitive) result!
One can also show that (for non-negative integers $a$ and $b$)
\begin{equation}
\left ( \partial^a \delta(x) \right ) \left ( \partial^b \delta (x)\right )
= 0,
\label{delsq4}
\end{equation}
because
\begin{equation}
\int_{H_{-\infty}}^{H_{\infty}} k^a (p-k)^b \, dk = 0.
\label{delsq5}
\end{equation}
Similarly one can show that, for non-negative integers $m$ and $a$,
the Fourier transform of $x^{-m} \partial^a
\delta(x)$ vanishes, and the Fourier transform of 
$x^m \partial^a \delta(x)$ is zero for $m>a$ and $i^{m+a} a(a-1)...(a-m+1)
p^{a-m}$ for $m \leq a$ respectively. 

%In doing quantum field theory in configuration space, we multiply
%operator-valued distributions of quantum fields.  Or, in a slightly
%different interpretation, we multiply
%singular functions such as Feynman propagators.  As we will see, the use of
%neutrix calculus allows one to put these products on a mathematically sound
%basis.  
Lastly let us generalize
the above discussion for products of two
1-dimensional Dirac delta functions to the
case of
(3 + 1)-dimensional Feynman propagators
\begin{eqnarray}
\Delta_{+} (x) &=& \int {d^4p \over (2 \pi)^4} {e^{ip \cdot x} \over
{p^2 + m^2 -i\epsilon}}\nonumber\\
&=& {1 \over 4 \pi} \delta (x^2) - {m \over 8 \pi  \sqrt{-x^2 - i\epsilon}}
H_1^{(2)} \! (m \sqrt{-x^2 - i\epsilon}),
\label{propag}
\end{eqnarray}
where $H^{(2)}$ is the Hankel function of the second kind
and we use the (+++-) metric.  The Fourier
transform of $\Delta_{+} (x) \times \Delta_{+} (x)$ (which appears in
certain quantum loop calculations, e.g., in Eq.~(\ref{phi3}) in the text)
is given by
\begin{eqnarray}
\int d^4x \, e^{-ip \cdot x} \, \Delta_{+} (x) \, \Delta_{+} (x)
&=& \int {d^4k \over (2 \pi)^4} \frac{1}{k^2 + m^2 - i\epsilon}
\frac{1}{(p - k)^2 + m^2 - i\epsilon}\nonumber\\
&=& \frac{i}{4 (2 \pi)^2} D - \frac{i}{4 (2 \pi)^2}
\int_0^1 dz \, \log \left (1 + {p^2 \over m^2} z (1 - z) \right),
\label{Delsq}
\end{eqnarray}
where $D$ is given by Eq.~(\ref{sigma4}).
But $D$ is logarithmically divergent.  Hence $\Delta_{+} (x) \times
\Delta_{+} (x)$ is not mathematically well defined.  On the other
hand, this product is mathematically meaningful in the 
Hadamard-van der Corput approach.  To wit,  $D = -1$ in
the neutralized version, hence $\Delta_{+} (x) \times \Delta_{+} (x) \sim
\delta^{(4)} (x) +$ regular part,
where $\delta^{(4)} (x)$ is the 4-dimensional Dirac delta function.

\end{document}